\documentclass[aps,prd,twocolumn,a4paper,floatfix,groupedaddress,preprintnumbers]{revtex4}
\usepackage{amsmath}
\usepackage{graphicx}

\begin{document}
\preprint{DFTT 37/2001}
\preprint{to be published in Phys.Rev.D}

\newcommand{\changedText}{}
\newcommand{\nbar}{\bar n}            
\newcommand{\Nbar}{\bar N}            
\newcommand{\nc}{{\bar n_c}}          
\newcommand{\pfrac}[3]{\left(\frac{\partial #1}{\partial #2}\right)_{\!\!#3}}
\newcommand{\avg}[1]{\langle #1 \rangle}  

\title{Thermodynamics of Clan Production}
\author{Alberto Giovannini}\author{Sergio Lupia}\author{Roberto Ugoccioni}
\affiliation{Dipartimento di Fisica Teorica, Universit\`a di Torino
and INFN - Sezione di Torino,
via P. Giuria 1, 10125 Torino, Italy}

\date{13 December 2001}

\begin{abstract}
Scenarios for particle production in the GeV and TeV regions are
reviewed.
The expected increase with c.m.\ energy of the average number of clans for
the soft component and the decrease for the semihard one indicate
possible classical and quantum behaviour of gluons, respectively.
Clan thermodynamics, discussed in the paper, appears as the natural
framework to deal with such phenomena.
\end{abstract}

\maketitle

\section{Introduction}
Phenomenological analysis of multiparticle production
in hadron-hadron collisions in the GeV region \cite{combo:prd,combo:eta}
revealed interesting substructures, i.e.\ soft and semihard  events,  
without and with minijets respectively, each class of events being described
by a negative binomial (Pascal) multiplicity distribution (NB MD) 
with different values of the parameters, the average
charged multiplicity $\nbar$ and $k = \nbar^2 / (D^2  - \nbar )$.
$D$ is here the dispersion. The weighted composition of the two MD's
leads  to the observed final charged particle MD. 
Clan structure analysis in terms of the average number of clans, $\Nbar$, and
the average number of particle per clan, $\nc$, allows to interpret nicely
the onset of above mentioned substructures in the GeV region.

By extrapolating  $\nbar$ and  $k$ behaviour to the TeV region, three possible
scenarios \cite{combo:prd,combo:eta}
for the semihard component have been investigated.
The first one assumes that KNO scaling is satisfied both in the soft 
and semihard component. This situation should be compared with what is
assumed in the other two scenarios  for the semihard component where KNO 
scaling is strongly violated or has a QCD inspired behaviour through the
c.m.\ energy dependence of the corresponding NB MD parameters.  
Since, at the present stage of QCD, calculations of MD's and
correlations in the GeV and TeV regions cannot be performed in a sound
way, we can only rely on QCD inspired extrapolations of the parameters.
The last two scenarios for the semihard component (the soft component
is taken to be the same in all three scenarios),  although more realistic
than the first one,  lead   to a decreasing average number of clans 
and to the corresponding increase of the average number of particles 
per clan as the c.m.\ energy increases. Since clans are
independently produced by assumption  it would be important to understand
the real meaning of their decrease for c.m.\ energies in the TeV region,
a fact which seems to widen the motivations at the origin of the first  
introduction of clan concept in high energy 
phenomenology \cite{AGLVH:1,AGLVH:4}.

In addition, it should be pointed out that  clan structure analysis can be 
generalised to the huge class of discrete infinitely divisible distributions
(IDD), to which NB(Pascal)MD belongs, and therefore any result obtained 
in the framework of clan structure analysis can be easily extended to the 
full class of IDD's.

In the present paper after a short introduction on the most appealing 
statistical theories of multiparticle production a new interpretation of 
$n$ charged particles multiplicity distribution, $p(n)$, for the class of IDD is 
proposed in terms of the canonical and grand-canonical partition functions.
Then the connection of this new interpretation for  the  description 
of the soft and semihard substructures   in  $hh$ collisions in terms of
the NB (Pascal) MD is examined.

\section{An alternative approach to multiparticle production}
One of the best known statistical approaches to multiparticle
dynamics is Feynman's fluid 
analogy \cite{Wilson:fluid,Bjorken:fluid}, where the
cross-section for the production of $n$ particles plays the role
of the partition function in the canonical ensemble, as it is an
integral over phase-space of the square of a matrix element which
plays the role of the Gibbs distribution, $e^{-H/k_B T}$. 
In this approach 
the volume is identified with the extension of phase
space and the fugacity $z$ with the dummy 
variable $u$ appearing in the definition of
the generating function $G$:
\begin{equation}
	G(u) \equiv \sum_n u^n p(n) .  \label{eq:gf}
\end{equation}
This identification is unsatisfactory because one
has to satisfy at the same time the definitions of the average number
of particles, from the grand-canonical ensemble:
\begin{equation}
	\avg{n} = z \frac{\partial \ln G}{\partial z}
\end{equation}
and from the definition of generating function:
\begin{equation}
	\avg{n} = \left.\frac{d G}{d u}\right|_{u=1} =
					\left.\frac{d \ln G}{d u}\right|_{u=1}  .
\end{equation}
The above formulae can be satisfied at the same time only in the limit of
zero chemical potential. 

Another approach was proposed by Scalapino and Sugar
\cite{ScalapinoSugar}:
they defined the probability amplitude to produce
a particle at rapidity $y$, denoted by $\Pi(y)$, as a random field
variable, then introduced a functional $F[\Pi]$ which played a role
analogous to the free energy for a system in thermal equilibrium.
One can then obtain the $n$-particle inclusive distribution
by averaging the product of the squares of the amplitudes,
$\Pi^2(y_1)\cdots\Pi^2(y_n)$, with a weight
given by $e^{-F[\Pi]}$.
Lacking the knowledge necessary to calculate $F[\Pi]$ from the underlying 
dynamics, the authors parametrised it (following Ginzburg and Landau)
in retaining the first three terms in a series expansion, then solved
the model in a few particular cases.
Remarkably, to leading order in the size of the allowed full rapidity range,
they obtain a generating function which has the form of an IDD.

More recent results obtained in the above mentioned frameworks,
concerning KNO scaling and phase transitions, can be found in 
\cite{Thermo:Biebl,Thermo:Uematsu,Thermo:Carruthers,%
Thermo:Antoniou:3,Thermo:Antoniou:2,Thermo:Hwa:1,Thermo:Hwa:2}.

Stimulated by these results we propose a new simplified approach to
the statistical theory of multiparticle production, heavily based on
IDD properties and valid for any chemical potential.

We denote with 
$Q_n(V,T)$ the partition function in the canonical ensemble
for a system with 
a fixed number $n$ of particles, volume $V$ and temperature $T$,
and with
${\cal Q}(z,V,T)$ the grand-canonical partition function
for a system with fugacity $z$, volume $V$ and temperature $T$; the
chemical potential $\mu$ is defined by $z = \exp(\mu/k_B T)$,
where $k_B$ is Boltzmann constant.

We recall the relation between the partition functions:
\begin{equation}
	{\cal Q}(z,V,T) = \sum_{n=0}^\infty  z^n  Q_n(V,T) .
\end{equation}
Accordingly, in a statistical mechanics framework,
the probability $p(n)$ of 
finding $n$ particles in the system is the following:
\begin{equation}
 	p(n) = \frac{z^n  Q_n(V,T)}{{\cal Q}(z,V,T)}	
			=	\frac{1}{n!} \frac{z^n}{\cal Q}
					\left.\frac{\partial^n {\cal Q}}{\partial z^n}\right|_{z=0}
				.		\label{eq:gcprob}
\end{equation}
Noticing that $Q_0(V,T) = 1$, we find immediately that the 
grand-canonical partition function is the inverse of the void probability
$p(0)$, i.e., of the probability to find no particles in the system:
\begin{equation}
	p(0) = \left[ {\cal Q}(z,V,T) \right]^{-1}  .    \label{eq:pzero}
\end{equation}
This result is very general: the void probability is the inverse
of the grand-canonical
partition function and all properties of the system can be
obtained from it \cite{Void}.

Consider now the wide class of power series distributions (PSD),
usually defined  as follows: 
\begin{equation}
	p(n) = \frac{ a_n b^n }{ \gamma(b) }	,			\label{eq:psd}
\end{equation}
with $a_n$ and $b$ free parameters, while proper
normalisation requires that
\begin{equation}
	\gamma(b) = \sum_{n=0}^\infty  a_n b^n .
\end{equation}
Notice that $a_0$ can always be chosen to be 1 (by redefining $\gamma(b)$
as $\gamma(b)/a_0$). Then one has for the void probability:
\begin{equation}
	p(0) = \frac{1}{\gamma(b)}  .
\end{equation}

Comparing Eq.s~(\ref{eq:gcprob}) to (\ref{eq:psd}),
our new approach is characterised by
the following correspondence:
\begin{align}
	z        &\longleftrightarrow b ,\nonumber\\
	Q_n      &\longleftrightarrow a_n ,  \label{eq:identify}\\
	{\cal Q} &\longleftrightarrow \gamma(b) = p(0)^{-1} .\nonumber
\end{align}

A very interesting property of this novel identification
is that $a_n$ is the canonical partition function for a system with a fixed
number of particles $n$, and in particular $a_1$ is the canonical partition
function for a system with 1 particle. This means that if we know the
multiplicity distribution of a thermodynamical system,  and cast
it into a PSD form, we can not only deduce the grand-partition function but
also identify the fugacity of the system and the canonical partition
function.
As an intriguing example of this correspondence, motivated
by the phenomenological analysis of multiparticle
production in the GeV region,
we will in the next section examine the NB(Pascal)MD.

\section{The negative binomial distribution}\label{sec:theNBD}
Any discrete infinitely divisible distribution (IDD)
can be written as a compound Poisson distribution (CPD)
\cite{Feller},
i.e., the number of clans $\Nbar$ can be defined
in such a way that the void probability is:
\begin{equation}
	p(0) = \exp( -\Nbar ) .		\label{eq:p0}
\end{equation}
Comparing with Eq.~(\ref{eq:pzero}), we notice that for any IDD
the average number of clans is the logarithm of the grand partition
function:
\begin{equation}
	\Nbar = \ln {\cal Q}  .    \label{eq:clans}
\end{equation}
All thermodynamical properties
can then be obtained by differentiating the average number of clans.
In particular, being for the grand canonical ensemble $PV=k_BT\ln{\cal Q}$,
we obtain the following
equation of state:
\begin{equation}
	PV = \Nbar k_B T  ;
\end{equation}
it tells us that our system behaves as an ideal gas of clans, an interpretation
which fits very nicely with the idea that clans are independent
objects, as implied by the definition of CPDs.

The NB(Pascal)MD, with parameters $\nbar$ and $k$:
\begin{equation}
	p(n) = \frac{k(k+1)\dots(k+n-1)}{n!} 
			\left( \frac{\nbar}{\nbar+k} \right)^n
			\left( \frac{k}{\nbar+k} \right)^k
\end{equation}
is an example of a PSD, with the following identification:
\begin{align}
	a_n &= \frac{k(k+1)\dots(k+n-1)}{n!} ,\nonumber\\
	b   &=  \frac{\nbar}{\nbar+k} , \label{eq:bnk}\\
	\gamma(b) &= (1-b)^{-k} .\nonumber
\end{align}
Furthermore, the NBMD also belongs to the class of discrete IDD,
the multiplicity distribution inside each clan being of logarithmic type.

We obtain therefore the following value for $\Nbar$:
\begin{equation}
	\Nbar = - k \ln (1-b) ;    \label{eq:Nbar=lnb}
\end{equation}
which also gives the the grand-canonical partition function,
applying eq.~(\ref{eq:clans}):
\begin{equation}
  {\cal Q} = (1-b)^{-k} .   \label{eq:Q=1^k}
\end{equation}

Comparing now with our proposed correspondence,
eq.~(\ref{eq:identify}),  we find that
$k$ is $a_1$, i.e., the canonical partition function 
for a system with 1 particle, and must therefore be function
of $V$ and $T$: $k=k(V,T)$;
we also find that $b$ is 
the fugacity of the system, i.e., $b = \exp(\mu/k_B T)$,
and it is a scaling function of $\nbar/k$, see eq.~(\ref{eq:bnk}).

We calculate now, using the standard thermodynamical relations,
the average number of particles in the system, $\avg{n}$,
which turns out to be equal to the $\nbar$ parameter of the NBMD:
\begin{equation}
	\begin{split}
	\avg{n} &= k_B T \pfrac{\Nbar}{\mu}{T,V} = 
	           - k_B T k \pfrac{\ln(1-b)}{\mu}{T,V} \\
			&= k_B T \frac{k}{1-b} \frac{b}{k_B T} = \frac{kb}{1-b}
			= \nbar ,
	\end{split}
\end{equation}
consistently with the above mentioned relation 
$b = \nbar/(\nbar+k)$.
We also obtain that the average number of particles per clan, $\nc$,
is a function only of the fugacity of the system:
\begin{equation}
	\nc = \frac{b}{(b-1) \ln(1-b)} .   \label{eq:nc=bonly}
\end{equation}
This result is very interesting because formally,
as already pointed out, $b$ is a scaling function of $\nbar/k$, and
experimentally $\nc$ is seen
to vary with the width of the rapidity interval at fixed c.m.\ energy:
if one were to identify (pseudo-)rapidity with volume, one would
expect that changing $\Delta\eta$ at fixed c.m.\ energy would 
imply changing the volume at constant temperature and fugacity
(intensive variables), thus keeping $\nc$ constant, contrary to 
observations. 
We must conclude that in the present approach we cannot identify
rapidity with volume as a simple thought would suggest but we should
allow volume to vary also with other physical quantities.

We now turn our attention to
the generating function (g.f.) for the multiplicity distribution,
defined in eq.~(\ref{eq:gf}).
In the general case illustrated by eq.~(\ref{eq:gcprob}) we easily
find
\begin{equation}
	G(u) = \sum_n \frac{u^n z^n Q_n}{\cal Q} = 
					\frac{{\cal Q}(uz,V,T)}{{\cal Q}(z,V,T)}
				= \frac{p(0)|_z}{p(0)|_{uz}} ,
\end{equation}
which is the ratio of the grand-canonical partition function 
(or of the void probability) at two different fugacities. 
This is a very general expression valid for any
system in the grand-canonical ensemble.

The g.f. for a CPD can always be written as
\begin{equation}
	G_{\text{CPD}}(u) = \exp \left[ \Nbar g(u) - \Nbar \right] ,
\end{equation}
where $g(u)$ is the g.f.\ for the multiplicity distribution within
each clan (it satisfies $g(0)=0$).

Because of eq.~(\ref{eq:clans}) we can write
for the class of CPDs:
\begin{equation}
	G(u) = e^{-\Nbar} {\cal Q}(uz,V,T) .
\end{equation}
However, $\ln {\cal Q}(uz,V,T) = \Nbar(uz,V,T)$, hence we obtain
\begin{equation}
	G(u) = \exp \left[ \Nbar(uz)-\Nbar(z) \right]
\end{equation}
which can be interpreted as a function of 
the \textit{difference} in the average numbers 
of clans for a system with fugacity $uz$
to that for a system at the actual fugacity $z$, 
keeping the same volume and temperature.
For the g.f.\ within one clan we further find:
\begin{equation}
	g(u) = \frac{\Nbar(uz)}{\Nbar(z)} 
				= \frac{\Omega(uz,V,T)}{\Omega(z,V,T)}
				= \frac{P(uz,V,T)}{P(z,V,T)} .
\end{equation}
Interestingly, this is the \textit{ratio} of the
average number of clans for two systems with
unequal fugacities.

In addition it is interesting to remark that, remembering that
parameter $k$ depends on $V$ and $T$, a complete thermodynamics can be
built in the just mentioned framework.
Its main quantities are listed in the following and 
explicitly calculated in the Appendix.

The equation of state is
\begin{equation}
	\frac{PV}{k_BT} = k \ln\left( 1+\frac{\avg{n}}{k} \right) .
\end{equation}
thus the average number of clans
can be expressed in terms of the thermodynamic potential $\Omega$:
\begin{equation}
	\Nbar = -\Omega / k_B T .
\end{equation}

The Helmholtz free energy
can be rewritten in a form symmetric in $\nbar$ and $k$:
\begin{equation}
  -\frac{A}{k_B T} = \avg{n} \ln \left(1+\frac{k}{\avg{n}}\right) + 
											k \ln\left(1+\frac{\avg{n}}{k}\right)  .
\end{equation}

The average internal energy is
\begin{equation}
  \frac{U}{k_B T} = \Nbar \pfrac{\ln k}{\ln T}{V}  .
\end{equation}

The entropy is
\begin{equation}
	\begin{split}
	S &= k_B\left\{ 
						-\frac{A}{k_B T} + T \pfrac{k}{T}{V} 
								\ln\left( 1+\frac{\avg{n}}{k} \right)
				\right\} .
	\end{split}  \label{eq:entropia}
\end{equation}
which coincides with $-A/T$ in the limit of $(\partial k/\partial T)_V
\to 0$, which gives also $U\to 0$.

In the next section we will focus our attention on clan thermodynamics
of final charged particle MD.

\section{Clan thermodynamics and the NB(Pascal)MD}

In this Section  an attempt is made in order to interpret 
\changedText
in the present approach a surprising finding in some  of the possible 
scenarios for hadron hadron collisions in the TeV 
region discussed in references \cite{combo:prd,combo:eta}, 
i.e. the unexpected decrease 
with c.m.\ energy   of the average number of clans for 
semihard events in scenarios 2 and 3. 

We are guided  by two considerations. 1) the occurrence of the NB(Pascal)MD
---as it is the case in the scenarios mentioned above both for semihard
and soft events--- is usually interpreted  as the result of a two step
process: to the independent production of clans during the first step,
it follows their  decay according to a logarithmic distribution, which
can be obtained  by a weighted average of geometric (Bose-Einstein)
distributions during the second step.
2) the validity of the generalised local parton hadron duality (GLPHD).

It should be pointed out that clan ancestors are independently produced
and Poissonianly distributed, by assumption, and a clan is, by definition,
a group of partons  of common ancestor; a clan consists of at least one 
parton, its ancestor. Each ancestor can be considered as  an independent 
intermediate gluon source. All correlations among generated partons are 
exhausted within each clan.

Clan ancestors can be produced either very early in the production process at
higher virtualities or later at lower virtualities .

In the first case, the ancestor's ``temperature'' 
(an unknown  function in this
approach of the average $p_T$ and the rapidity) is expected to be
higher: this 
expectation, together with the  lack of  mutual correlations among  
ancestors, emphasises their overall quasi-classical behaviour:
ancestor production  in this case is
competitive with the increase of gluon population within each clan.
This situation is qualitatively closer to that expected at hadron
level for soft events and semihard events in scenario 1.

In the second case ancestors are produced later, at lower virtualities:
their ``temperature'' should also be lower, with even 
``colder'' generated gluons.
Their virtuality is lower. Accordingly,   quantum effects should be expected 
to be enhanced in events sharing these properties:
new produced gluons prefer to stay together with other relatives within
each clan than to become ancestor and initiate a new clan. $k$ parameter
is in this case lower  and closer to that of a Bose-Einstein distribution,
which occurs for $k = 1$. These remarks are consistent
with the interpretation of $1/k$ (see \cite{AGLVH:4}) 
as a measure of aggregation
of partons into clans: it corresponds to the ratio of the probability
to have two gluons (particles at hadron level) in the same clan  over
the probability to have  two gluons in two separate clans,  i.e., to
smaller $k$ parameter corresponds an  higher aggregation among produced
gluons into clans.
In addition, being $1/k$ linked  to the integral of two parton
rapidity correlations  via second order factorial cumulants, the decrease 
of  $k$ implies stronger two parton correlations. In conclusion generated
gluons prefer  to stay together  than  to stay far apart,
higher parton density regions are generated, the probability 
to create a new gluon is enhanced (a typical quantum 
effect), clans become more populated  and their average number is reduced.
This situation is closer qualitatively to that expected at hadron level for
semihard events in scenarios 2 and 3 of references 
\cite{combo:prd,combo:eta}.

The just mentioned considerations and 
Eq.s (\ref{eq:Nbar=lnb}), (\ref{eq:Q=1^k}) and (\ref{eq:nc=bonly})
of Section \ref{sec:theNBD}
fully outline in the present approach the importance of the behaviour of 
the fugacity variable $b$.

The remaining question is how clan thermodynamics results at parton
level can be extended to final particles through the hadronization
mechanism. 
A possible answer to this question comes from generalised local
parton-hadron duality (GLPHD) \cite{AGLVH:2} 
which says that all inclusive distributions are proportional at the two
levels of investigation:
\begin{equation}
	Q_{n,\text{hadrons}}(y_1,\dots,y_n) =
		\rho^n Q_{n,\text{partons}}(y_1,\dots,y_n) ,
\end{equation}
which corresponds for NBMD parameters to
\begin{equation}
 k_{\text{hadron}} = k_{\text{parton}} ,\qquad
	\nbar_{\text{hadron}} = \rho \nbar_{\text{parton}} .
\end{equation}
GLPHD can be applied separately to soft and semihard components thus solving
our problem.
In particular in this framework minijets production is related to the
existence of regions of high gluon densities and final particle
production should be sensitive to the mentioned quantum effects,
by increasing two particle correlations and BE effects.

Motivated by these considerations, the behaviour of parameter $b$ as a
function of c.m.\ energy, as well as the $b$ dependence of $\Nbar$ and
$\nc$ have been explored as a suggestive example in the different
above mentioned
scenarios phenomenologically described in terms of NBMD's.
In addition in view of their simple connections with $\Nbar$ and $\nc$,
the probability of having no particles in the event, $p(0)$,
and the void scaling function, ${\cal V}(\nbar/k)$, have been studied
as a function of fugacity $b$. It turns out that the analysis in terms
of $p(0)$ and ${\cal V}(\nbar/k)$ variables confirms the main result of the
new approach, i.e., that the reduction of the average number of clans
with the increase of c.m.\ energy is a quantum effect.

\begin{figure*}
  \begin{center}
  \mbox{\includegraphics[width=0.92\textwidth]{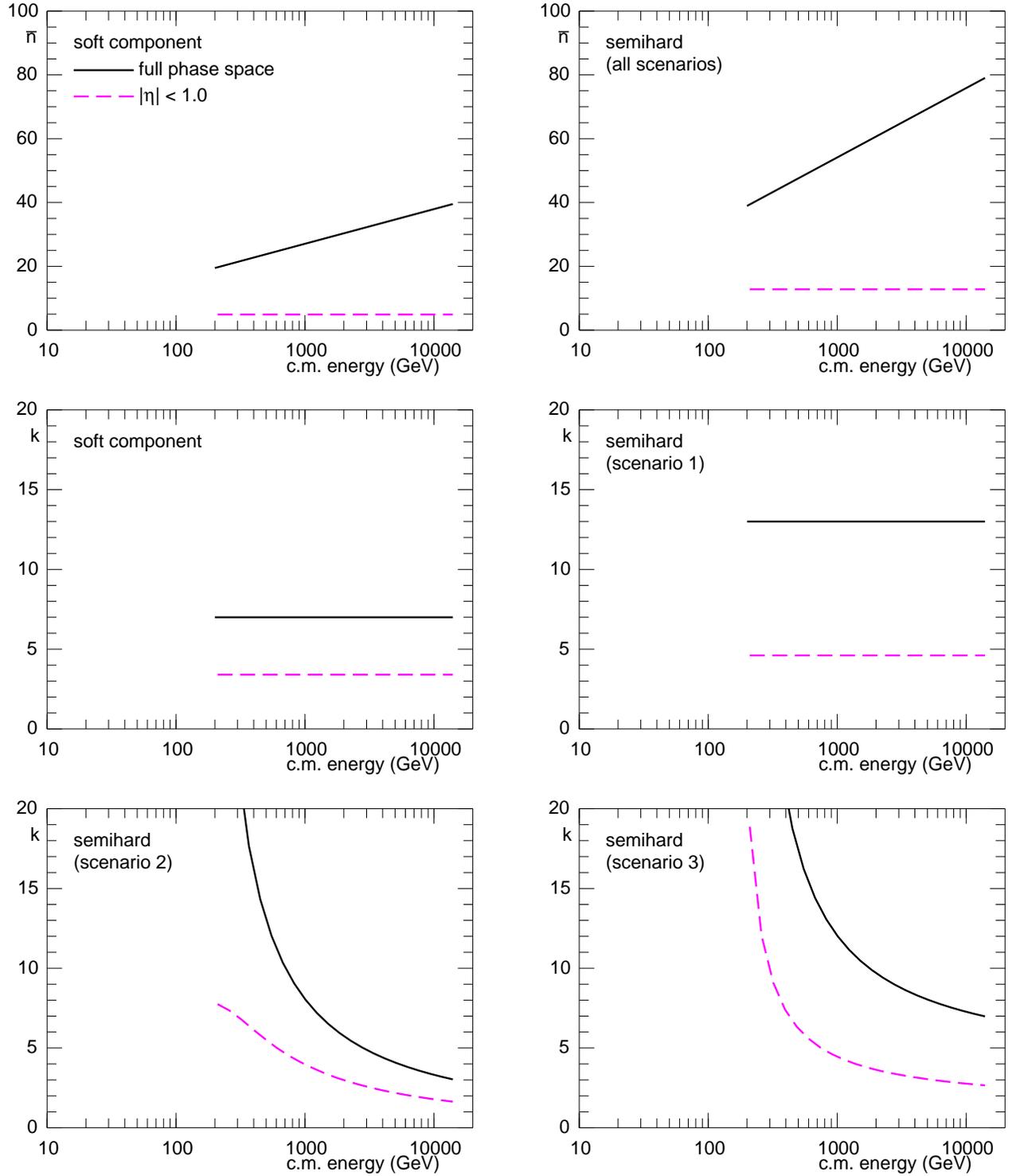}}
  \end{center}
  \caption{C.m.\ energy dependence of standard NBMD parameters 
$\nbar$ and $k$ in the three  scenarios described in the
introduction. The top two panes show the behaviour of $\nbar$
(it is the same in all scenarios for both the semihard and the soft
component). The lower four panes show the $k$ parameter for
the soft component (equal in all scenarios) and the semihard
one in the three scenarios.
}\label{fig:nk}
  \end{figure*}

\begin{figure*}
  \begin{center}
  \mbox{\includegraphics[width=0.92\textwidth]{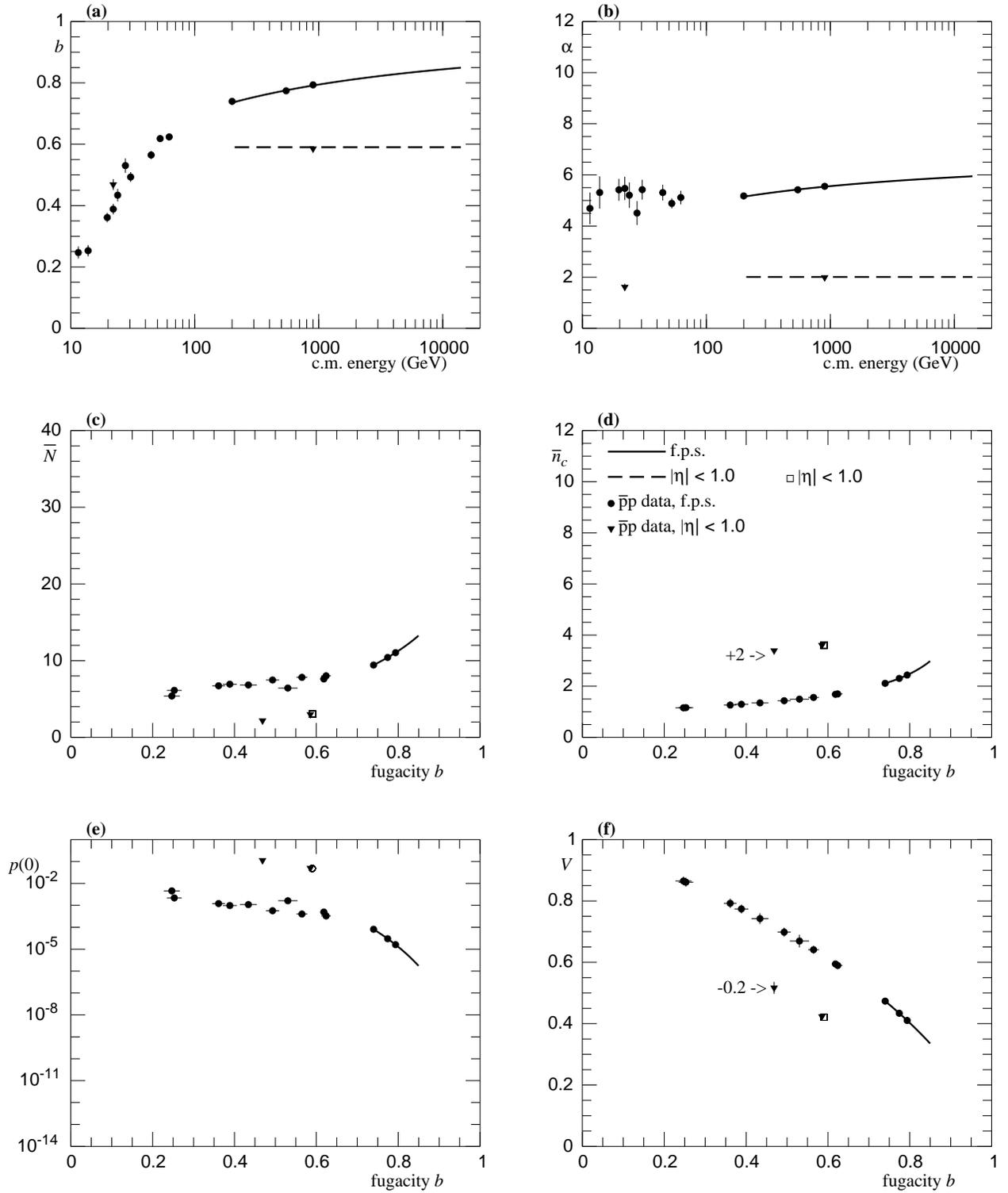}}
  \end{center}
  \caption{Results for the soft component (equal in all scenarios).
In this figure, the lines and the open points 
show the results from our extrapolations:
the solid line refers to full phase space, the dashed lines
and open square to the interval 
$|\eta|<1$. The last point on each line correspond to a c.m.\ energy
of 14 TeV.
The solid circles show full phase space data from ISR and UA5, the 
solid triangles refer to UA5 data 
in the interval $|\eta|<1$.
\textbf{(a)} fugacity $b$ as a function of c.m.\ energy;
\textbf{(b)} $\alpha$ parameter as a function of c.m.\ energy;
\textbf{(c)} average number of clans vs fugacity; notice that 
this is also a plot
of the grand partition function in logarithmic scale, 
since $\Nbar = \log{\cal Q}$, Eq.~(\ref{eq:clans});
\textbf{(d)} average number of particles per clans vs fugacity; here
for clarity 
the curves and the data for 
$|\eta|<1$ are shifted up by 2 units;
\textbf{(e)} void probability vs fugacity;
\textbf{(f)} void scaling function ${\cal V}$ vs fugacity; also here
curves and the data for 
$|\eta|<1$ are shifted down by 0.2 units.
}\label{fig:soft}
  \end{figure*}

\begin{figure*}
  \begin{center}
  \mbox{\includegraphics[width=0.92\textwidth]{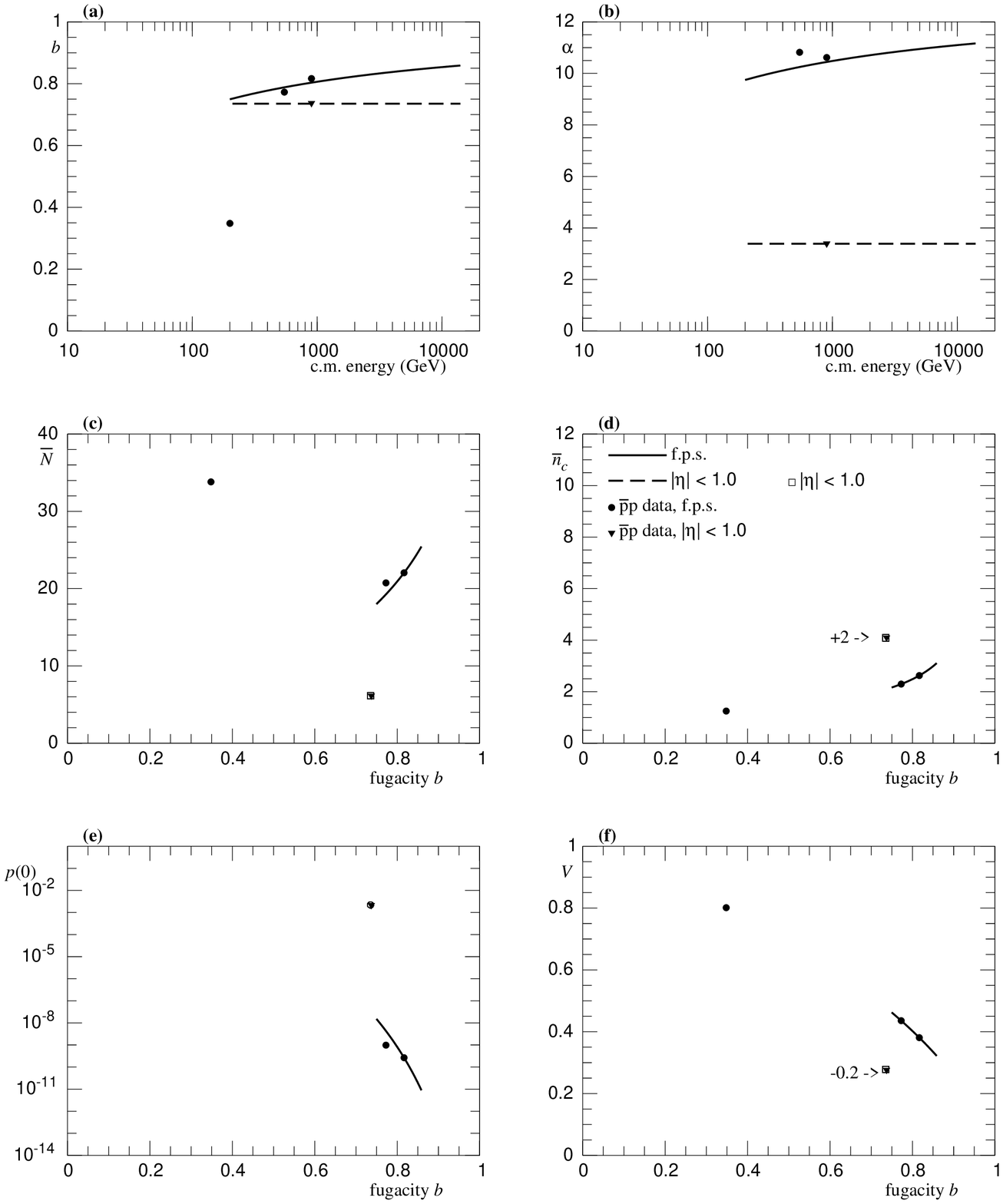}}
  \end{center}
  \caption{Same as figure~\ref{fig:soft}, but the semihard
component in scenario 1 is shown.}\label{fig:semi1}
  \end{figure*}

\begin{figure*}
  \begin{center}
  \mbox{\includegraphics[width=0.92\textwidth]{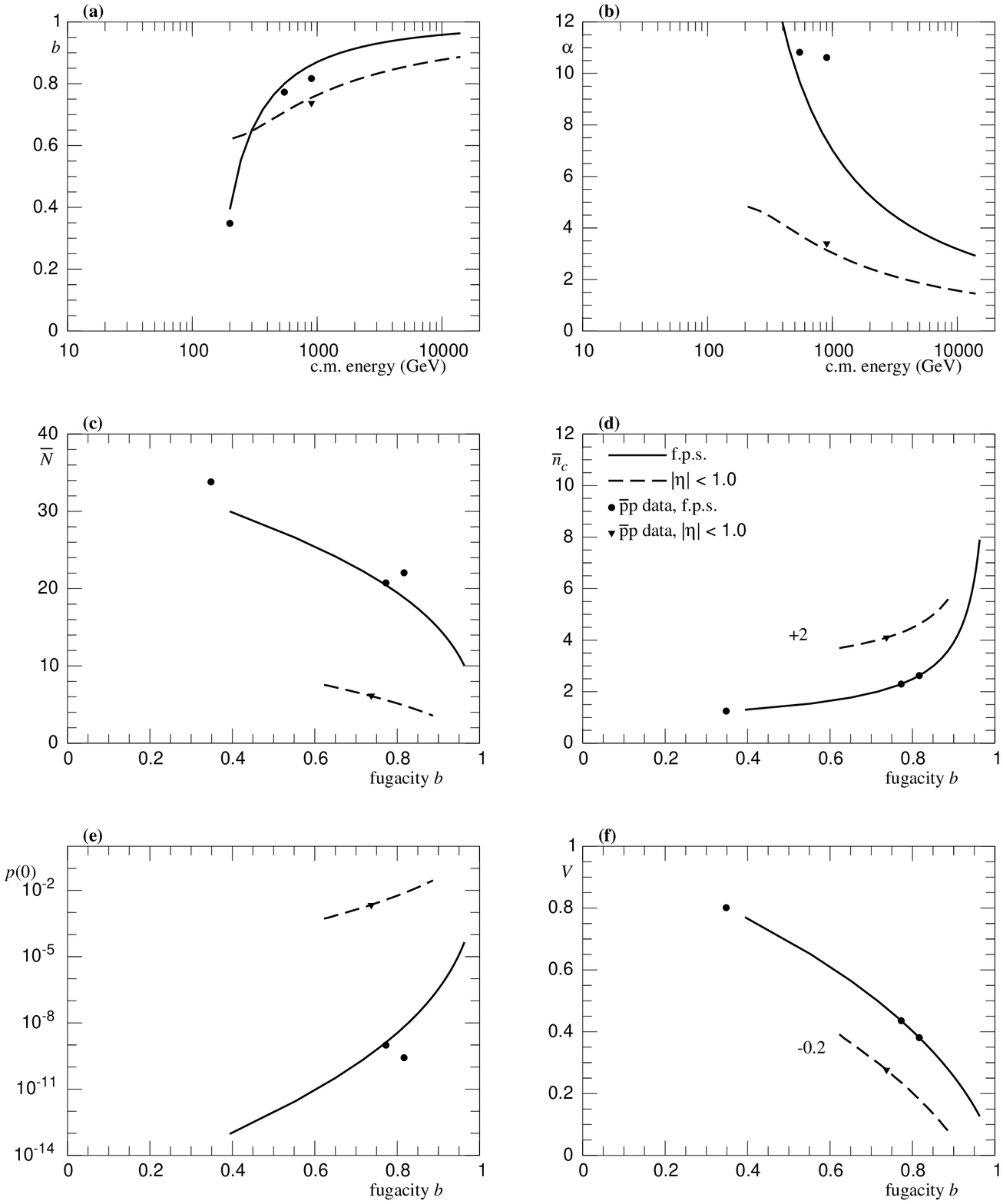}}
  \end{center}
  \caption{Same as figure~\ref{fig:soft}, but the semihard
component in scenario 2 is shown.}\label{fig:semi2}
  \end{figure*}

\begin{figure*}
  \begin{center}
  \mbox{\includegraphics[width=0.92\textwidth]{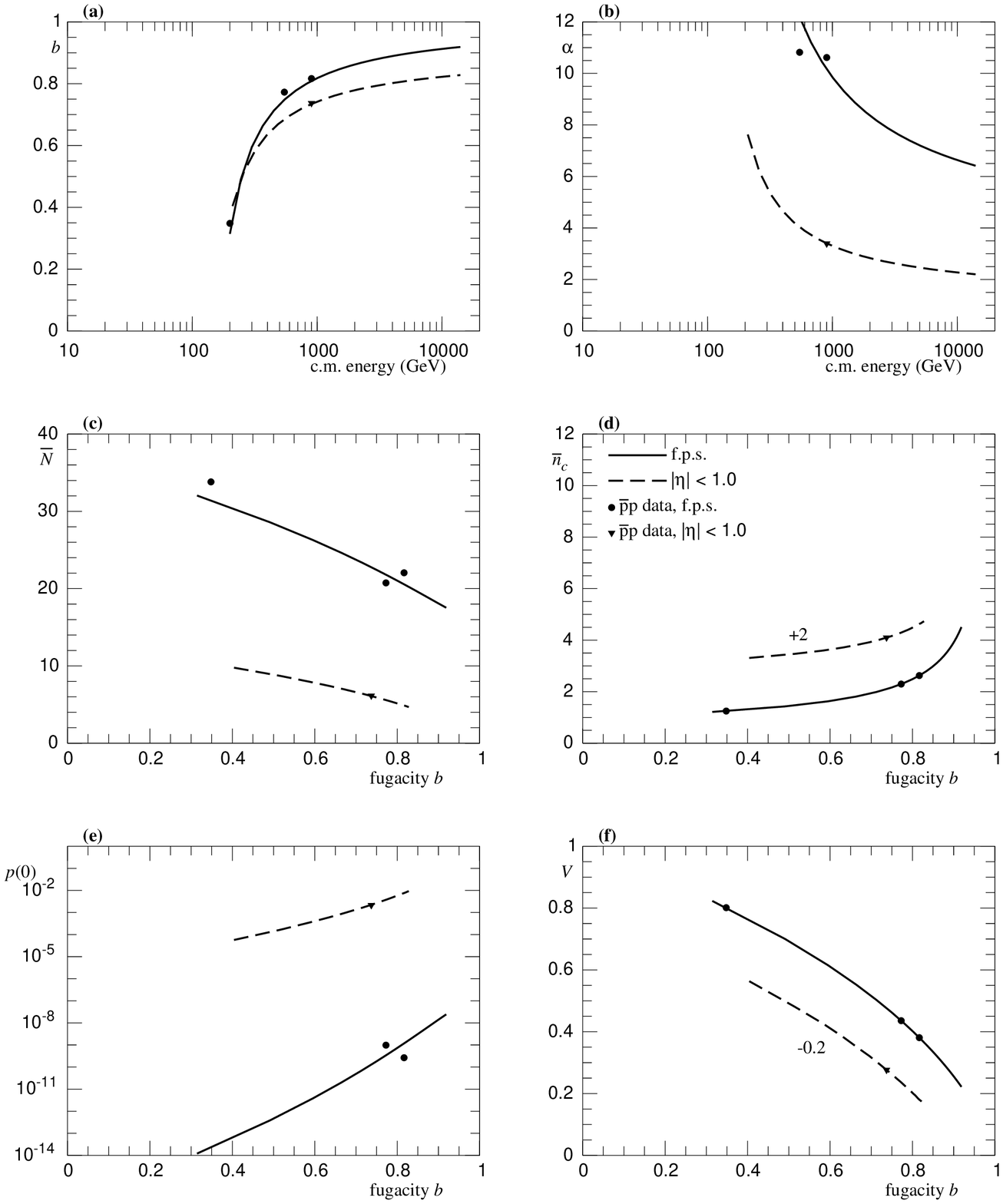}}
  \end{center}
  \caption{Same as figure~\ref{fig:soft}, but the semihard
component in scenario 3 is shown.}\label{fig:semi3}
  \end{figure*}

We proceed now to discuss the thermodynamical behaviour of multiparticle
production according to the scenarios described in the introduction
and fully characterised by the c.m.\ energy dependence shown in
Fig.~\ref{fig:nk}.
We start with the soft component thermodynamical behaviour since
it is assumed to be the same in all above mentioned scenarios
(see Fig.~\ref{fig:soft}).

We observe that fugacity $b$ is growing  very fast  from 0.25 to 0.75 
with c.m.\ 
energies below 100 GeV and then smoothly varying from 0.75 up to 0.9 at
14 TeV. The explanation of this behaviour  will be given in terms of the
following three known possible interpretation of the occurrence on 
NB(Pascal)MD in high energy physics phenomenology, which  in terms of
parameters $\nbar$ and $k$  are:

\begin{subequations}
\begin{equation}
\textit{a)}\qquad\quad
  \frac{(n+1)p(n+1)}{p(n)} = \alpha + \beta n  \label{eq:x.a}
\end{equation}
where $\alpha= k \nbar/( \nbar + k )$  and $\beta = \nbar/ (\nbar+ k)$.
Notice that for $\alpha = \beta$,  i.e.\ for $k = 1$, the MD $p(n)$ becomes a 
Bose-Einstein distribution, for $\beta= 0$, i.e.\ $k\to\infty$,
a Poissonian distribution (the Poissonian limit) and for $\alpha=0$ a
logarithmic distribution, which can be expressed as the superposition
of Bose-Einstein distributions.

\begin{equation}
\textit{b)}\qquad\quad
	\Nbar = k\ln\left(1+\frac{\nbar}{k}\right) ,\qquad
	\nc = \frac{\nbar}{k\ln(1+\nbar/k)}        \label{eq:x.b}
\end{equation}
where $\Nbar$ is the average number of clans and $\nc$ the
average number of particles per clan.

\begin{equation}
\textit{c)}\qquad\quad
	p(0) = \left(\frac{k}{\nbar+k}\right)^{k} , \qquad
	{\cal V} = \frac{k}{\nbar}\ln  \left(1+\frac{\nbar}{k}\right)
		          \label{eq:x.c}
\end{equation}
where $p(0)$ is the probability of generating zero charged particles and
${\cal V}(\nbar/k)$ is the void scaling function; the occurrence of scaling
in the product of the first two moments $\nbar$ and $1/k$  indicates two 
particle correlation dominance for hierarchical systems,
and  the distance from point one (Poisson limit ${\cal V}(0)=1$) on the scaling 
function is larger,  more numerous and correlated are the particles.  
\end{subequations}

It should be pointed out  that  $\beta$ parameter in
Eq.~(\ref{eq:x.a}) 
coincides with the fugacity $b$ discussed above  
in our thermodynamical approach and therefore $b$, 
like ${\cal V}(\nbar/k)$, is a scaling function of $\nbar/k$, and 
$\alpha$ parameter  corresponds to the average charged  multiplicity, $\nbar$,
for  a classical system ($k\to\infty$).

In this sense the  relative behaviour of $\beta=b$ 
\changedText
and $\alpha=kb$ as the c.m.\ energy increases  in view of 
Eq.\ (\ref{eq:x.a}) and the 
discussion at the beginning of 
this Section, can be considered  an indication of the relative importance 
of a behaviour closer to a quantum  one, i.e.\ harder, with respect to a 
behaviour closer to a quasi-classical, i.e.\ softer, for a class of events.  
A very slow increase of $b$ with c.m.\ energy and an almost constant behaviour
of $\alpha=kb$ is the main  characteristic of the class of soft events as
shown in Figures \ref{fig:soft}a and \ref{fig:soft}b.

This fact is confirmed by inspection of Fig.~\ref{fig:soft}c and d, where it is
shown that $\Nbar$ is a very slow growing  function of the fugacity of 
the system throughout the ISR region and below that region ($\approx 7$), and
then a quickly growing function of the same variable in the GeV region
up to 14 in the TeV region (14 TeV); $\nc$ as a function of the fugacity
has a similar behaviour  from $\approx 1.5$ to $\approx 3$.

Accordingly, the probability of creating zero charged particles, $p(0)$, is
decreasing throughout the same regions from $10^{-2}$ to $10^{-5}$ (for $b=0.9$
at 14 TeV c.m.\ energy); in addition (Fig.~\ref{fig:soft}f) the void
scaling function 
${\cal V}(\nbar/ k)$ turns out to populate for larger values of fugacity 
variable sections of the curve  far from the Poissonian limit (${\cal V}(0)=1$) 
showing a clear increase   of two particle correlations in this region
as expected for a hierarchical system.

It should be noticed that in the soft  scenario  (no minijets) constant
$k$ parameter behaviour in rapidity intervals as requested by KNO scaling
in the GeV and TeV regions implies that also the other variables remain
constant in the same regions. 

In scenario 1, as already  pointed out, the semihard component is assumed  to
have a very similar behaviour to the soft one: KNO scaling is satisfied
and minor changes  in the general trend of the variables both in full
phase space and in (pseudo)rapidity intervals  are straightforward
consequences of the smaller constant $k$ parameter value suggested by
NB fits for the semihard component in the GeV and extrapolated 
to the TeV region (Fig.~\ref{fig:semi1}).

Coming to the second scenario the assumption of strong KNO scaling
violation for the semihard component (an extreme point of view with
respect to that of scenario 1)  implies a completely new panorama with
dramatic changes. Fugacity $b$ (Fig.~\ref{fig:semi2}a) is growing very 
fast from 0.4 
at 200 GeV c.m.\ energy up to  0.96 at 14 TeV almost saturating the
maximum allowed value, which is one, and $\alpha$ parameter 
(Fig.~\ref{fig:semi2}b)
is decreasing very rapidly from $\approx 16$ at 200 GeV to $\approx 3$
at 14 TeV.

The combined information contained in the two figures leads to the same 
conclusion, i.e.\ the proposed semihard scenario behaviour is much closer to a
\changedText
quantum one than the soft scenario favouring the production of regions
of higher  particle density.   This interpretation  is
confirmed by studying general trends of the other variables  as  a function
of fugacity $b$. The  average number of clans $\Nbar$ is decreasing  in full
phase space  from $\approx 30$ ($b \approx 0.35$) to $\approx 10$  at
14 TeV ($b \approx 0.96$) and the 
average number per clan, $\nbar_c$ is increasing from $\approx 1$ to
$\approx 8$ in the
same interval. Accordingly the probability of zero particle production  is
increasing from  $\approx 10^{- 13}$ to $\approx 10^{-4}$,    
i.e.\  gap probability
is increasing  with fugacity  and c.m.\ energy; in parallel void
scaling function $V(\nbar/k)$ is populating sections  with much higher 
$b$ values than in scenario 1 corresponding to regions  much farther
from the Poissonian limit.  One interesting point concerns smaller 
(pseudo)rapidity intervals (say $|\eta| < 1$): the general trend is that
 $\Nbar$ is lower than in full phase space as  are corresponding $\nc$
values, thus suggesting the onset of regions with higher particle densities
and lower temperatures. The probability of generating zero particles,
in view of the higher densities, is therefore much higher and the void
probability is far from the Poisson limit.

Scenario  3 is a QCD inspired scenario  for the semihard component:
it assumes  for parameter $k$ a QCD behaviour 
(see Fig.~\ref{fig:semi3}). This scenario gives a
panorama  for our variables which is intermediate between the two
extremes, 1 and 2. Fugacity $b$  is increasing very fast with c.m.\ energy as
in scenario 2, but $\alpha$ parameter  has a sweeter trend  (it is $\approx 6$ 
at 14 TeV) indicating stronger independent production; this 
fact is clearly shown in Fig~\ref{fig:semi3}c,d where   $\Nbar$ is larger and 
$\nc$ smaller than in scenario 2. Differences in $p(0)$ and ${\cal V}(\nbar/k)$
behaviours  in full phase space as well as in (pseudo)rapidity intervals
with respect to scenario 2 are all consequences of the just mentioned 
remarks.

\section{Conclusions}
Clan thermodynamics has been investigated in order to explain the
decrease with c.m.\ energy of $\Nbar$ for the semihard component and
its increase for the soft one in the most realistic scenarios of
multiparticle production in the GeV and TeV regions in $hh$
collisions.
It turns out that these two behaviours for clans point out
structures closer to classical or quantum properties of gluons.
A thermodynamical approach to multiparticle production was constructed on
this basis.
Results were determined in the framework of NB(Pascal)MD applied separately in
the two components and can be extended to any infinitely divisible
distribution. 

\appendix*
\section{}
The main quantities of clan thermodynamics
are explicitly calculated in the following.

The Helmholtz free energy is
\begin{equation}
	A = \avg{n}\mu - PV = \frac{kb}{1-b} \mu - k_B T \Nbar .
\end{equation}

The average internal energy is
\begin{equation}
	\begin{split}
	U &= k_B T^2 \pfrac{\Nbar}{T}{b,V}\\ &=
			- k_B T^2 \pfrac{k}{T}{V} \ln(1-b) \\
		&= k_B T^2 \pfrac{k}{T}{V} \ln\left( 1+\frac{\avg{n}}{k} \right)\\ &=
			k_B T^2  \frac{\Nbar}{k} \pfrac{k}{T}{V} .
	\end{split}
\end{equation}

\begin{widetext}
The entropy is
\begin{equation}
	\begin{split}
	S &= \frac{U-A}{T} = k_B \Nbar - \avg{n} \frac{\mu}{T} + k_B T
				\frac{\Nbar}{k} \pfrac{k}{T}{V} =\\
		&= k_B \Nbar + 
				k_B \avg{n} \ln\left( 1+\frac{k}{\avg{n}} \right) +
				k_B T \pfrac{k}{T}{V} \ln\left( 1+\frac{\avg{n}}{k} \right) =\\
		&= k_B\left\{  
						\left[k + T \pfrac{k}{T}{V} \right]
									\ln\left( 1+\frac{\avg{n}}{k} \right) + 
						\avg{n} \ln\left( 1+\frac{k}{\avg{n}} \right) 
				\right\}   .
	\end{split}
\end{equation}

The specific heat at constant volume is
\begin{equation}
	\begin{split}
		C_v &= 2 k_B T \pfrac{k}{T}{V} \ln\left( 1+\frac {\avg{n}} k \right) +
						k_B T^2 \pfrac{^2k}{T^2}{V} \ln\left( 1+\frac {\avg{n}} k \right)
						\\	&\quad + k_B T^2 \pfrac{k}{T}{V} \frac{1}{1+\avg{n}/k}
						\left(-\frac{\avg{n}}{k^2}\right) \pfrac{k}{T}{V}\\
			&= 2\frac{U}{T} + 
					k_B T^2 \pfrac{^2k}{T^2}{V} \ln\left( 1+\frac {\avg{n}} k \right) 
					- k_B T^2 {\pfrac{k}{T}{V}}^2 \frac{\avg{n}}{k(k+\avg{n})} .
	\end{split}
\end{equation}
\end{widetext}

\begin{acknowledgments}
We would like to dedicate this paper to Enrico Fermi on the occasion
of the 100th anniversary of his birth, remembering his famous
paper on the thermodynamical model of multiparticle dynamics
\cite{Fermi}.
\end{acknowledgments}


\end{document}